\documentclass[11pt]{article}
\title{\bf  Massless Thirring model in canonical quantization scheme}
\author{S.E. Korenblit, V.V. Semenov}
\begin{document}
\maketitle
\begin{abstract}
It is shown that the exact solubility of the massless Thirring model in the canonical 
quantization scheme originates from the intrinsic hidden linearizability of its 
Heisenberg equations in the method of dynamical mappings. The corresponding role of 
inequivalent representations of free massless Dirac field and appearance of Schwinger  
terms are elucidated.
\end{abstract}

\section{Introduction}

Despite a considerable age the two-dimensional Thirring model
\cite{thi}--\cite{s_w} is still remained as important
touchstone for non-perturbative methods of quantum field theory
\cite{Leut}--\cite{man} revealing new features both in the
well-known \cite{nak}--\cite{fab-iva} and in newly obtained
solutions \cite{fuj-hir-hom-tak}. At the same time the methods
of integration of such two-dimensional models provide a key for
understanding some non-linear theories of higher dimension
\cite{fab-iva}. In particular the Thirring model turns out to
be a two-dimensional analog of the well-known
Nambu-Jona-Lasinio model \cite{fab-iva}, \cite{fuj-hir-hom-tak}
and together with the Schwinger model provides an important
example of using the well-known bosonization procedure (BP)
\cite{col}--\cite{blot}.

In the present work the BP for Thirring model is considered as a special case of
dynamical mapping (DM) \cite{mtu}, \cite{green_1}, what for Schwinger model was
previously done in Greenberg's works \cite{green}. In the framework of canonical
quantization scheme \cite{hep} the DM method consists in the construction of Heisenberg
field (HF) $\Psi(x)$ as a solution of Heisenberg equations of motion (HEq) in the
form of Haag expansion built on normal products \cite{gldj} of free ``physical'' fields
$\psi(x)$, whose representation space accords with unknown {\it a priori}
physical states of the given field theory \cite{mtu}. The DM
$\Psi(x)\stackrel{\rm w}{=}\Upsilon[\psi(x)]$, being generally
speaking a weak equality, implies the choice of appropriate
initial conditions for the HEq. For example \cite{blot}, \cite{mtu}, when
both sets of fields are complete, irreducible and coincide
asymptotically as $t\to -\infty$, the HF will tend in a weak
sense to appropriate asymptotic physical field $\psi_{in}(x)$:
$\lim\limits_{t\to-\infty}\Psi(x^1,t)\stackrel{\rm w}{=}\Upsilon[\psi_{in}(x^1,-\infty)]$.
However the (asymptotic) completeness and irreducibility are not true in
the presence of bound states \cite{mtu}, \cite{green_1}. In particular for the exactly
solvable two-dimensional models of Thirring and Schwinger \cite{fab-iva}, \cite{blot}
the physical asymptotic states of propagated physical particles have nothing
to do with massless free Dirac asymptotic fields (confinement).

As was shown in the works \cite{vklt}--\cite{ks} it is more
convenient generally to make DM onto the ``Schr\"odinger''
physical field $\psi_s(x)$, associated with the HF at $t\to 0$:
$\lim\limits_{t\to 0}\Psi(x^1,t)\stackrel{\rm
w}{=}\Upsilon[\psi_s(x^1,0)]$, which is a generalization
\cite{kt}, \cite{ks} of the well-known interaction
representation and is closely related to the procedure of
canonical quantization \cite{blot}, \cite{hep}. In this representation the
time-dependent coefficient functions of DM \cite{vklt},
\cite{kt} contain all the information about bound states and
scattering, and exactly solvable Federbush model \cite{wai} leads to
the exactly linearizable HEq \cite{ks}.

The present paper shows that HEq of the Thirring model admits a similar linearization and
that the choice of free massless (pseudo-) scalar fields as the physical ones is a
consequence of reducibility of the massless Dirac field \cite{blot} in the space of
these fields.
The problem of Schwinger terms in the currents commutator \cite{Leut}, being closely
related to BP \cite{nak}--\cite{blot}, also finds here a natural solution \cite{ks} in
fact borrowed from QED \cite{sok}, where it is also sufficient to define this commutator
only for the free fields in corresponding ``interaction representation''.

Definition of the model in canonical quantization scheme is
given in the next section. Then the linearization procedure
with corresponding definition of Heisenberg currents is
advocated. The bosonization rules that we need for the free
fields only are discussed in section 4 with the appropriate
choice of (pseudo-) scalar fields. That all is used in section
5 for direct integration of HEq with chosen initial condition.
The final remarks are made in section 6.

\section{Thirring model}

Following to the canonical quantization procedure \cite{hep} we
start with the formal Hamiltonian of the Thirring model
\cite{thi}, which in two-dimensional space-time\footnote{Here:
$x^\mu=\left(x^0,x^1\right)$; $x^0=t$; $\hbar=c=1$;
$\partial_\mu=\left(\partial_0,\partial_1\right)$; for
$g^{\mu\nu}$: $g^{00}=-g^{11}=1$; for $\epsilon^{\mu\nu}$:
$\epsilon^{01}=-\epsilon^{10}=1$;
$\overline{\Psi}(x)=\Psi^\dagger(x)\gamma^0$;
$\gamma^0=\sigma_1$, $\gamma^1=-i\sigma_2$,
$\gamma^5=\gamma^0\gamma^1=\sigma_3$, $\gamma^\mu\gamma^5=
-\epsilon^{\mu\nu}\gamma_\nu$, where $\sigma_i$ -- Pauli
matrices, and $I$ -- unit matrix; $x^\xi = x^0 + \xi x^1$,
$2\partial_\xi =2{\partial}/{\partial
x^\xi}=\partial_0+\xi\partial_1$, $P^1=-i\partial_1$; summation
over repeated $\xi=\pm $ is nowhere implied.} defines a Fermi
self-interaction, with fixed (and further unrenormalizable)
dimensionless coupling constant $g$, for spinor field with spin
$1/2$ and zero mass:
\begin{eqnarray}
&&
H[\Psi]=H_{0[\Psi]}(x^0)+H_{I[\Psi]}(x^0),
\label{K_0} \\
&&
H_{I[\Psi]}(x^0)=\frac{g}{2}\int\limits_{-\infty}^\infty dx^1
J_{(\Psi)\,\mu}(x) J_{(\Psi)}^\mu (x),
\label{vbnmei1} \\
&&
H_{0[\Psi]}(x^0)=\int\limits_{-\infty}^\infty
dx^1\Psi^\dagger(x)E(P^1)\Psi(x), \quad E(P^1)=\gamma^5P^1,
\label{K_3}
\end{eqnarray}
satisfying the equal-time canonical anticommutation relations:
\begin{eqnarray}
&&
\left\{\Psi_\xi(x), \Psi_{\xi'}^\dagger (y) \right\} \bigr|_{x^0= y^0} =
\delta_{\xi, \xi'} \delta \left(x^1 - y^1\right),
\label{vbnmei2} \\
&&
\left\{\Psi_\xi(x),\Psi_{\xi'} (y)\right\} \bigr|_{x^0=y^0}=0,
\nonumber \\
&&
\left\{\Psi_\xi(x),\Psi^{\#}_{\xi'}(y)\right\} \bigr|_{(x-y)^2<0}=0,\;\,\mbox{ with: }\; 
\Psi^{\#}_\xi(y)=\Psi_\xi(y),\;\Psi^{\dagger}_\xi(y). 
\label{vbnmei4}
\end{eqnarray}
Here indices $\xi, \xi' = \pm $, as well as for the $x^\xi$, enumerate the components 
of HF by the rule:
\begin{eqnarray}
\Psi (x) = \left(\begin{array}{cc}
\Psi_1(x) \\ \Psi_2(x) \end{array}\right)=
\left(\begin{array}{cc}
\Psi_{+}(x) \\ \Psi_{-}(x)\end{array}\right),
\label{vbnmei6}
\end{eqnarray}
and the vector current $J_{(\Psi)}^\mu(x)$, together with the axial current
$J_{(\Psi)}^{5\mu}(x)$, for $\mu,\nu = 0,1$, is their yet formal local bilinear
functional of the form:
\begin{eqnarray}
&&
J_{(\Psi)}^\mu (x)\longmapsto \overline{\Psi}(x)\gamma^\mu \Psi(x),
\label{vbnmei5} \\
&&
J_{(\Psi)}^{5\mu}(x)\longmapsto\overline{\Psi}(x)\gamma^\mu\gamma^5\Psi(x)=
-\epsilon^{\mu\nu}J_{(\Psi)\nu}(x),
\nonumber
\end{eqnarray}
which due to (\ref{K_0})--(\ref{vbnmei6}) formally appears also in the canonical 
equations of motion\footnote{Contribution to (\ref{45bn6l4}) due to non-commutativity of 
$J_{(\Psi)}^\nu(x)$ and $\Psi(x)$ is formally proportional to 
$\delta(0)\gamma^0\gamma_\nu\gamma^0\gamma^\nu=0$.} \cite{s_w}--\cite{wai}:
\begin{eqnarray}
&&
i\partial_0\Psi(x)=\left[\Psi(x), H[\Psi]\,\right]=
\left[E(P^1)+g\gamma^0\gamma_\nu J_{(\Psi)}^\nu(x)\right]\Psi(x),
\label{45bn6l4} \\
&&
\mbox{or: }\;
2\partial_\xi \Psi_\xi(x)=-igJ^{-\xi}_{(\Psi)}(x)\Psi_\xi(x), \quad  \xi=\pm ,
\label{45bn6l5}
\end{eqnarray}
-- for each $\xi$-component of the field (\ref{vbnmei6}) that
formally are also related to the corresponding current
components as:
\begin{eqnarray}
J_{(\Psi)}^\xi (x)=J_{(\Psi)}^0 (x)+\xi J_{(\Psi)}^1(x)\longmapsto
2\Psi_\xi^\dagger(x)\Psi_\xi (x), \quad \xi=\pm .
\label{vbnmei8}
\end{eqnarray}
The correct definitions of these formal operator products will be discussed hereinafter.

\section{Linearization of the Heisenberg equation}

An immediate consequence of the field equations of motion
(\ref{45bn6l4}), (\ref{45bn6l5}) are the local conservation
laws \cite{s_w}--\cite{wai} for the currents (\ref{vbnmei5}),
(\ref{vbnmei8}):
\begin{eqnarray}
\partial_\mu J_{(\Psi)}^\mu(x)=0,\quad
\partial_\mu J_{(\Psi)}^{5\mu}(x)=-\epsilon_{\mu\nu}\partial^\mu J^\nu_{(\Psi)}(x)=0,
\;\mbox{ or: }\; \partial_\xi J_{(\Psi)}^\xi(x)=0,\quad \xi=\pm ,
\label{CC_1} 
\end{eqnarray}
that fully determine their dynamics as a free one \cite{Leut}, \cite{d_f_z}.
Therefore it is not surprising that by means of the same equations of motion
(\ref{45bn6l4}), (\ref{45bn6l5}), as well as by means of the anti-commutation
relations (\ref{vbnmei2}) for HF, it is a simple matter to show
\cite{ks} that:
\begin{eqnarray}
i\partial_0\gamma^0\gamma_\nu J_{(\Psi)}^\nu(x)-
\left[\gamma^0\gamma_\nu J_{(\Psi)}^\nu (x), H_{0[\Psi]}(x^0)\right]
= iI\,\partial_\mu J_{(\Psi)}^\mu(x)+
i\gamma^5\,\epsilon_{\mu\nu}\partial^\mu J^\nu_{(\Psi)}(x)\equiv 0,
\label{bnemti4}
\end{eqnarray}
where the first term on the r.h.s. of equality (\ref{bnemti4})
comes evidently from the left terms with $\nu=0$, while the
second term on the r.h.s. comes from the left terms with
$\nu=1$. The canonical equation of motion for this operator of
``total current'' in Eq. (\ref{45bn6l4}), containing of course
its commutator with the full Hamiltonian $H[\Psi]$ given by
Eqs. (\ref{K_0})--(\ref{K_3}), recasts then to the following equation:
\begin{eqnarray}
i\partial_0\gamma^0\gamma_\nu J_{(\Psi)}^\nu(x)-
\left[\gamma^0\gamma_\nu J_{(\Psi)}^\nu (x), H_{0[\Psi]}(x^0)\right]
=\left[\gamma^0\gamma_\nu J_{(\Psi)}^\nu (x), H_{I[\Psi]}(x^0)\right] = 0,
\label{bnemti5}
\end{eqnarray}
which thus can not contain a contribution from the commutator
with the interaction Hamiltonian $H_{I[\Psi]}(x^0)$ given by Eq. (\ref{vbnmei1}).
Hence, as well as for the Federbush model \cite{ks}, a non-zero contribution of
Schwinger terms in HEq (\ref{bnemti5}) would be premature, because, due to Eq.
(\ref{bnemti4}), it leads to violation of the current conservation laws (\ref{CC_1}).

On the one hand, within the framework of canonical quantization procedure \cite{hep},
the vanishing of expressions (\ref{bnemti4}), (\ref{bnemti5}) means that temporal
evolution of this ``total current'' is governed by a free Hamiltonian
$H_{0[\chi]}\left(x^0\right)$ of the same form (\ref{K_3}) quadratic on some kind of
free massless trial physical Dirac fields $\chi(x)$, furnished by the same
anti-commutation relations and by the same conservation laws for corresponding currents
$J_{(\chi)}^\nu (x)$, $J_{(\chi)}^{5\nu}(x)$, defined formally by
Eqs. (\ref{vbnmei2})--(\ref{vbnmei5}),  (\ref{vbnmei8}), (\ref{CC_1}) with
$\Psi(x)\mapsto\chi(x)$:
\begin{eqnarray}
i\partial_0\gamma^0\gamma_\nu J_{(\chi)}^\nu (x)-
\left[\gamma^0\gamma_\nu J_{(\chi)}^\nu (x), H_{0[\chi]}(x^0)\right]=
iI\,\partial_\mu J_{(\chi)}^\mu(x)+
i\gamma^5\,\epsilon_{\mu\nu}\partial^\mu J^\nu_{(\chi)}(x)=0.
\label{bnemti6}
\end{eqnarray}
On the other hand, the Heisenberg current operators appearing in (\ref{bnemti4}),
(\ref{bnemti5}) acquire precise operator meaning -- with non-vanishing Schwinger
term -- only after the choice of the representation space \cite{Leut}, \cite{hep},
\cite{Vlad} for anticommutation relations (\ref{vbnmei2}), (\ref{vbnmei4}) and
subsequent reduction in this representation to the normal-ordered form by means of
renormalization, for example, via point-splitting and subtraction of the vacuum
expectation value \cite{blot}:
\begin{eqnarray}
&&\!\!\!\!\!\!\!\!\!\!\!\!\!\!\!\!\!\!
J^0_{(\Psi)} (x) \longmapsto
\lim\limits_{\widetilde{\varepsilon} \rightarrow 0}
\widehat{J}^0_{(\Psi)}(x;\widetilde{\varepsilon})=\widehat{J}^0_{(\Psi)}(x),
\quad J^1_{(\Psi)} (x) \longmapsto
\lim\limits_{\varepsilon \rightarrow 0}
\widehat{J}^1_{(\Psi)}(x;\varepsilon)=\widehat{J}^1_{(\Psi)}(x),
\label{bos-111}\\
&&\!\!\!\!\!\!\!\!\!\!\!\!\!\!\!\!\!\!
\mbox{where at first: }\; \widetilde{\varepsilon}^0 = \varepsilon^1\rightarrow 0,\;
\mbox{ with fixed: }\; \widetilde{\varepsilon}^1 =\varepsilon^0,\;\;\;
\varepsilon^2=-\widetilde{\varepsilon}^2>0,
\label{K_E} \\
&&\!\!\!\!\!\!\!\!\!\!\!\!\!\!\!\!\!\!
\mbox{for: }\;
\widehat{J}^\nu_{(\Psi)}(x;a)=
Z^{-1}_{(\Psi)}(a)\left[\overline{\Psi}(x + a)\gamma^\nu \Psi (x)-
\langle 0|\overline{\Psi}(x + a)\gamma^\nu\Psi(x)|0\rangle\right],
\label{K_Z}
\end{eqnarray}
and accordingly for the components (\ref{vbnmei8}). The
renormalization ``constant'' $Z_{(\Psi)}(a)$ is defined below
in (\ref{K_Za}). The definition of renormalized current
(\ref{bos-111})--(\ref{K_Z}) used here corresponds to the
well-known Schwinger prescription \cite{sok} specified in the
work \cite{ot} and, unlike Johnson definition \cite{Jon},
directly depends on the representation choice via the vacuum
expectation value \cite{blot} in Eq. (\ref{K_Z}) like the very
meaning of Schwinger term \cite{Leut}, \cite{fab-iva}. One can
show \cite{ot} that for the massless case these different
current definitions lead to coincident expressions only for the
free Dirac fields (cf. Eqs. (\ref{nweyi19}) and (\ref{J_L})
below).

The comments given above jointly with the foregoing arguments deduced from Eq.
(\ref{CC_1})--(\ref{bnemti6}) allow to identify in HEq (\ref{45bn6l4}), at least in a
weak sense, the Heisenberg operator of ``total current'', defined by Eqs.
(\ref{vbnmei5}), (\ref{bnemti4}), with that operator, defined by Eqs. (\ref{vbnmei5}),
(\ref{bnemti6}) for the free massless trial physical Dirac fields $\chi(x)$ and
renormalized in the sense of normal form (\ref{bos-111})--(\ref{K_Z}) up to an unknown
yet constant $\beta$:
\begin{eqnarray}
&&
\gamma^0\gamma_\nu J_{(\Psi)}^\nu(x) \stackrel{\rm w}{\longmapsto}
\frac{\beta}{2\sqrt{\pi}}\gamma^0\gamma_\nu\widehat{J}_{(\chi)}^\nu(x),
\label{ns94m61} \\
&&
\widehat{J}_{(\chi)}^\nu(x)=
\lim\limits_{\varepsilon,(\widetilde{\varepsilon})\rightarrow 0}
\widehat{J}_{(\chi)}^\nu\left(x;\varepsilon (\widetilde{\varepsilon})\right)\,
\equiv\, :J_{(\chi)}^\nu(x):\,.
\label{ns94m62}
\end{eqnarray}
Here for $Z_{(\chi)}(a)=1$ the symbol $:\ldots:$ means the usual normal form \cite{gldj}
with respect to free field $\chi(x)$. This identification leads to
linearization of both Eqs. (\ref{45bn6l4}), (\ref{45bn6l5}) in the representation of
these trial fields $\chi(x)$. Of course, the Eq. (\ref{45bn6l4}) is linearized with
respect to $x^0$, while the Eq. (\ref{45bn6l5}) -- with respect to $x^\xi$. However,
the latter equation is the preference of two-dimensional world with initial condition
being far from evidence. Whereas the former equation admits the above-mentioned
in the Introduction physically reasonable initial condition at $x^0=0$.
Unlike \cite{Leut}, \cite{blot}, \cite{ks}, this initial condition does not fix here
the constant $\beta$, which will be defined dynamically in subsequent sections.

\section{Bosonization and scalar fields}

As was shown in \cite{ks} such kind of linearization of HEq for the Federbush model
directly leads to its solution in the form of DM
$\Psi(x)=\Upsilon[\psi_{1}(x), \psi_{2}(x)]$ onto the free massive Dirac fields
$\psi_{1,2}(x)$ with different non-zero masses $m_{1,2}$. Unlike the massive one, the
components $\chi_\xi(x)$ of two-dimensional free massless field become completely
decoupled, $\partial_\xi \chi_\xi(x)=0$. As a consequence, this field turns out to be
essentially non-uniquely defined or reducible and equipped by many inequivalent
representations both in the spaces of a free massless (pseudo-) scalar field \cite{blot}
($\phi(x)$), $\varphi(x)$ and massive scalar field \cite{raja} $\phi_m(x)$. Because the
DM is physically meaningful only onto the complete, irreducible sets of fields:
$\Psi(x)=\Upsilon[\varphi(x),\phi(x)]$, or $\Psi(x)=\Upsilon[\phi_m(x)]$, or
$\Psi(x)=\Upsilon[\psi_M(x)]$, -- for the phase with spontaneously broken chiral
symmetry \cite{fab-iva}, \cite{fuj-hir-hom-tak}, further we consider here only
the first possibility.

The corresponding BP allows to operate with functionals of
boson fields instead of fermion operators and forms a powerful
tool for obtaining non-perturbative solutions in various
two-dimensional models \cite{nak}, \cite{fab-iva}, \cite{blot},
\cite{ks}. Its use also simplifies integration of the
linearized HEq (\ref{45bn6l4}).

Being a formal consequence of the current conservation conditions (\ref{CC_1}) only,
the bosonization rules have, generally speaking, the sense of weak equalities only for
the current operator in the normal-ordered form (\ref{bos-111})--(\ref{K_Z}), that
already implies a choice of certain representations of (anti-) commutation relations
(\ref{vbnmei2}) and (\ref{K_8_1}) below.
However, for the free massless fields $\chi(x)$, $\varphi(x)$, $\phi(x)$, this choice is
carried out automatically. This, due to the linearization condition
(\ref{ns94m61}), (\ref{ns94m62}), becomes enough for our purposes, since for the free
fields these relationships appear as operator equalities \cite{blot}:
\begin{eqnarray}
\widehat{J}_{(\chi)}^\mu(x)=\frac{1}{\sqrt{\pi}}\,\partial^\mu\varphi(x)=
-\,\frac{1}{\sqrt{\pi}}\,\epsilon^{\mu\nu}\partial_\nu \phi (x), \quad
\widehat{J}_{(\chi)}^{-\xi} (x) = 
\frac{2}{\sqrt{\pi}}\,\partial_{\xi} \varphi^{\xi}\left(x^{\xi}\right)\,.
\label{nweyi19}
\end{eqnarray}
Here, unlike \cite{nak}, the free massless scalar field $\varphi(x)$,
$\partial_\mu\partial^\mu\varphi(x)=0$, and pseudoscalar field $\phi(x)$,
$\partial_\mu\partial^\mu\phi(x)=0$, are mutually dual and coupled by symmetric
integral relations:
\begin{eqnarray}
&&
\left. \begin{array}{c}\phi(x) \\
\varphi(x)
\end{array}\right\}
=-\frac{1}{2}\int\limits_{-\infty}^\infty dy^1
\varepsilon \left(x^1-y^1\right)\partial_0
\left\{\begin{array}{c}\varphi\left(y^1,x^0\right), \\
\phi\left(y^1,x^0\right), \end{array}\right.
\label{K_5}
\end{eqnarray}
where the step function $\varepsilon(x^1)=1$, for $x^1>0$, $\varepsilon(x^1)=-1$, 
for $x^1<0$, $\varepsilon(0)=0$, and the corresponding charges for these fields have the 
form similar to \cite{fab-iva}, \cite{blot}:
\begin{eqnarray}
&&
\left. \begin{array}{c} O \\
\overline{O} \end{array}\right\}
=\int\limits_{-\infty}^\infty dy^1 \partial_0
\left\{\begin{array}{c}\varphi\left(y^1,x^0\right) \\
\phi\left(y^1,x^0\right)\end{array}\right\}= 
\left\{\begin{array}{c}\phi(-\infty, x^0)-\phi(\infty,x^0). \\
\varphi(-\infty,x^0)-\varphi(\infty,x^0). \end{array}\right.
\label{K_O} 
\end{eqnarray}
Right and left fields $\varphi^{\xi}\left(x^{\xi}\right)$ and their charges
${\cal Q}^\xi$ are defined by linear combinations \cite{blot}:
\begin{eqnarray}
\varphi^\xi\left(x^\xi\right)=
\frac{1}{2}\left[\varphi(x)-\xi\phi(x)\right],\qquad 
{\cal Q}^\xi=\frac{1}{2}\left[O-\xi\overline{O}\right]=
\pm 2\varphi^\xi\left(x^0\pm\infty\right),
\label{K_7}
\end{eqnarray}
for $\xi=\pm $. All commutation relations \cite{nak,fab-iva_2,blot} for the fields 
$\varphi(x)$, $\phi(x)$, $\varphi^\xi\left(x^\xi\right)$, and ${\cal Q}^\xi$:
\begin{eqnarray}
&&
\left[\varphi(x),\partial_0 \varphi (y)\right] \bigr|_{x^0=y^0}=
\left[\phi(x),\partial_0\phi(y)\right] \bigr|_{x^0=y^0}=i\delta(x^1-y^1),
\label{K_8_1} \\
&&
\left[\varphi(x), \varphi (y)\right]=
\left[\phi(x),\phi(y)\right]=
-\,\frac{i}{2}\,\varepsilon(x^0-y^0)\,\theta\left((x-y)^2\right),
\label{K_8} \\
&&
\left[\varphi^\xi\left(s\right),\varphi^{\xi'}\left(\tau\right)\right]=
-\frac{i}{4}\varepsilon(s - \tau)\delta_{\xi, \xi'}, \quad
\left[\varphi^\xi(s),{\cal Q}^{\xi'}\right]=\frac{i}{2}\delta_{\xi,\xi'},
\label{K_9}
\end{eqnarray}
are reproduced by commutators of their frequency parts and corresponding charges
\cite{d_f_z,nak,fab-iva}:
\begin{eqnarray}
&&
\left[\varphi^{\xi(\pm)}(s),\varphi^{\xi'(\mp)}(\tau)\right]=
\mp\frac{1}{4\pi}\ln\biggl(i\kappa\Bigl\{\pm(s-\tau)-i0\Bigl\}\biggl)
\delta_{\xi, \xi'},
\label{nblaie16} \\
&&
\left[\varphi^{\xi(\pm)}(s), {\cal Q}^{\xi'(\mp)}\right]=
\frac{i}{4}\delta_{\xi, \xi'}, \quad
\left[{\cal Q}^{\xi(\pm)}, {\cal Q}^{\xi'(\mp)}\right]=
\pm\frac{1}{4}\delta_{\xi, \xi'},
\label{nblaie19}
\end{eqnarray}
defined here by the creation/annihilation operators $c^\dagger(k^1),\,c(k^1)$ of the 
pseudoscalar field $\phi(x)$: 
${\cal P}c\left(k^1\right){\cal P}^{-1}=-c\left(-k^1\right)$, with 
$[c\left(k^1\right),c^\dagger\left(q^1\right)]=4\pi k^0\delta\left(k^1-q^1\right)$, and 
$k^0\equiv|k^1|$, as: 
\begin{eqnarray}
&&\!\!\!\!\!\!\!\!\!\!\!\!\!\!\!\!\!\!
\varphi^{\xi(+)}(s)= 
-\,\frac{\xi}{2\pi}\!\int\limits_{-\infty}^\infty \!\frac{d k^1}{2 k^0}
\theta \left(-\xi k^1\right)c\left(k^1\right) e^{-i k^0 s}, \quad 
\varphi^{\xi(-)}(s)=\left[\varphi^{\xi(+)}(s)\right]^\dagger,
\label{ph_pm} \\
&&\!\!\!\!\!\!\!\!\!\!\!\!\!\!\!\!\!\!
{\cal Q}^{\xi(+)}=\lim_{L\rightarrow\infty}\frac{iL\xi}{4\sqrt{\pi}}\!
\int\limits_{-\infty}^\infty\! d k^1 \theta\left(-\xi k^1\right) c\left(k^1\right) 
e^{-ik^0 x^0} e^{-\left(k^1 L/2\right)^2}, \quad 
{\cal Q}^{\xi(-)}=\left[{\cal Q}^{\xi(+)}\right]^\dagger. 
\label{Q_pm} 
\end{eqnarray}
According to \cite{fab-iva_2}, the invariance under the parity transformation 
${\cal P}\{\ldots\}{\cal P}^{-1}$ for generating functional of a free massless 
pseudoscalar field, unlike the scalar field theory, leads to its well definiteness and 
the gauge invariance also under field's shift by arbitrary constant. 
According to \cite{blot}, in such a well-defined space of bosonic fields 
(\ref{K_5})--(\ref{Q_pm}) one can construct the variety of different inequivalent 
representations of solutions of the Dirac equation for massless free trial field,
$\partial_\xi\chi_\xi(x)=0$, in the form of local normal-ordered exponentials of left
and right boson fields $\varphi^\xi(x^\xi)$ and their charges ${\cal Q}^\xi$ (\ref{K_7}), 
(\ref{K_9}). Let us choose the most simple of them \cite{blot}, which leads to the 
bosonization relations (\ref{nweyi19}) for the currents (\ref{bos-111})--(\ref{K_Z})
of trial fields $\chi(x)$ with $Z_{(\chi)}(a)=1$:
\begin{eqnarray}
&&
\chi_\xi\left(x\right)=
\chi_\xi\left(x^{-\xi}\right)= {\cal N}_\varphi\left\{\exp\left(- i \sqrt{\pi}
\left[2\varphi^{- \xi} \left(x^{-\xi}\right)+
\frac{\xi}{2}{\cal Q}^\xi\right]\right)\right\} u_\xi,
\label{nblaie12} \\
&&
u_\xi =\sqrt{\frac{\kappa}{2 \pi}}e^{-\pi/32}e^{-i\pi\xi/4}.
\nonumber
\end{eqnarray}
The infrared regularization parameter $\kappa$ from (\ref{nblaie16}) can subsequently 
tend to zero \cite{blot} or remain to be fixed, $\kappa\mapsto M$, \cite{fab-iva}, 
depending on the phase of the model under consideration. 

\section{Integration of the Heisenberg equation}

For the chosen representation (\ref{nweyi19})--(\ref{nblaie16})
the operator product in the linearized by means of (\ref{ns94m61}),
(\ref{ns94m62}) HEq (\ref{45bn6l4}) or (\ref{45bn6l5}) is
naturally redefined into the normal-ordered form \cite{blot}
with respect to the fields $\varphi^\xi(x^\xi)$:
\begin{eqnarray}
\partial_0 \Psi_\xi (x)=\left(-\xi \partial_1-i\frac{\beta g}{2\sqrt{\pi}}
\widehat{J}_{(\chi)}^{-\xi(-)} (x)\right) \Psi_\xi (x)-
\Psi_\xi(x)\left(i\frac{\beta g}{2\sqrt{\pi}}\widehat{J}_{(\chi)}^{-\xi(+)}(x)\right).
\label{nweyi13}
\end{eqnarray}
The famous expression for the derivative of function $F\left(x^1\right)$ in terms of the
operator $P^1$: $-i\partial_1 F(x^1)=\left[P^1,F(x^1)\right]$, and its finite-shift
equivalent: $e^{i a P^1}F(x^1)e^{- iaP^1}=F(x^1+a)$, allows to transcribe the equation
(\ref{nweyi13}) for $x^0=t$, $\Psi_\xi(x)\longleftrightarrow Y(t)$, as follows:
\begin{eqnarray}
\frac {d}{dt}Y(t) = A(t)Y(t)-Y(t)B(t),
\label{nweyi15}
\end{eqnarray}
and to obtain then its formal solution in the form of
time-ordered exponentials:
\begin{eqnarray}
Y(t) = {T}_A \left\{\exp\left(\int\limits_0^t d\tau
A(\tau)\right)\right\}Y(0)
\left[{T}_B \left\{\exp\left(\int\limits_0^t d\tau
B(\tau)\right) \right\}\right]^{-1},
\label{nweyi16}
\end{eqnarray}
that are immediately replaced here by the usual ones, recasting the solution already
into the normal form:
\begin{eqnarray}
\Psi_\xi (x)= e^{C^{\xi(-)}(x)}\Psi_\xi\left(x^1-\xi x^0, 0\right) e^{C^{\xi(+)}(x)},
\label{nweyi21}
\end{eqnarray}
where operator bosonization (\ref{nweyi19}) of the vector current of trial field 
$\chi(x)$ (\ref{nblaie12}) gives:
\begin{eqnarray}
&&
C^{\xi(\pm)} (x)=-i\frac{\beta g}{2\sqrt{\pi}}\int\limits_0^{x^0}d y^0
\widehat{J}_{(\chi)}^{-\xi(\pm)}\left(x^1+\xi y^0 -\xi x^0, y^0\right)=
\label{nweyi22}  \\
&&
 -i\frac{\beta g}{2\pi}\left[\varphi^{(\pm)}\left(x^1,x^0\right)-
\varphi^{(\pm)}\left(x^1-\xi x^0,0\right)\right]
= -i\frac{\beta g}{2\pi}\left[\varphi^{\xi(\pm)}\left(x^\xi\right)-
\varphi^{\xi(\pm)}\left(-x^{-\xi}\right)\right].
\nonumber
\end{eqnarray}
Remarkably, that the completely unknown ``initial'' HF
$\Psi_\xi(x^1-\xi x^0,0)=\lambda_\xi(x^{-\xi})$ appears here also as a solution of
free massless Dirac equation, $\partial_\xi\lambda_\xi(x^{-\xi})= 0$,
but certainly unitarily inequivalent to the free field $\chi(x)$ (\ref{nblaie12}).
The expressions  (\ref{nweyi21}), (\ref{nweyi22}) suggest to choose it also in the
normal-ordered form with respect to the field $\varphi$, using appropriate
``bosonic canonical transformation'' of this field with parameters
$\overline{\alpha}=2\sqrt{\pi}\cosh\eta$, $\overline{\beta}=2\sqrt{\pi}\sinh\eta$, 
obeying $\overline{\alpha}^2-\overline{\beta}^2 = 4\pi$, which is generated by the
operator $F_\eta$ (for $y^0=x^0$) in the form:
\begin{eqnarray}
&&\!\!\!\!\!\!\!\!\!\!\!\!\!\!\!\!\!\!\!\!
U^{-1}_\eta\varphi\left(x\right)U_\eta=\omega(x)\equiv
\omega^\xi\left(x^\xi\right)+\omega^{-\xi}\left(x^{-\xi}\right)=
\frac{1}{2\sqrt{\pi}}\Bigl[\overline{\alpha}\varphi(x^1,x^0)+
\overline{\beta}\varphi(x^1,-x^0)\Bigr], 
\label{om_U} \\
&&\!\!\!\!\!\!\!\!\!\!\!\!\!\!\!\!\!\!\!\!
U^{-1}_\eta\varphi^\xi\left(x^\xi\right)U_\eta=\omega^\xi\left(x^\xi\right)=
\frac{1}{2\sqrt{\pi}}\left[\overline{\alpha}\varphi^\xi \left(x^\xi\right)+
\overline{\beta}\varphi^{-\xi}\left(-x^\xi\right)\right], \quad U_\eta =\exp F_\eta,
\label{intro_007} \\
&&\!\!\!\!\!\!\!\!\!\!\!\!\!\!\!\!\!\!\!\!
U^{- 1}_\eta {\cal Q}^\xi U_\eta= {\cal W}^\xi= 
\frac{1}{2\sqrt{\pi}}\left[\overline{\alpha}{\cal Q}^\xi-
\overline{\beta}{\cal Q}^{-\xi}\right],\;\mbox{ with: }\;
\left[\varphi^{\xi(\pm)}(s),F_\eta\right]=\eta\,\varphi^{-\xi(\mp)}(-s), 
\label{intro_008} \\
&&\!\!\!\!\!\!\!\!\!\!\!\!\!\!\!\!\!\!\!\!
\mbox{for: }\;
F_\eta = 2i\eta\int\limits_{-\infty}^\infty d y^1 \varphi^\xi\left(y^\xi\right)
\partial_0\varphi^{-\xi}\left(-y^\xi\right)=
2i\eta\int\limits_{-\infty}^\infty d y^1 \omega^\xi\left(y^\xi\right)
\partial_0\omega^{-\xi}\left(-y^\xi\right),
\label{intro_009}
\end{eqnarray}
-- does not depend at all on $\xi$ and $y^0$, and ($\Lambda$ is ultraviolet 
cut-off \cite{fab-iva}): 
\begin{eqnarray}
&&\!\!\!\!\!\!\!\!\!\!\!\!\!\!\!\!\!\!\!\!
U^{-1}_\eta\chi_\xi\left(x^{- \xi}\right) U_\eta=
\lambda_\xi\left(x^{-\xi}\right)=
{\cal N}_\varphi \left\{\exp\left(-i\sqrt{\pi}\left[
2\omega^{-\xi}\left(x^{-\xi}\right)+
\frac{\xi}{2}{\cal W}^\xi\right]\right)\right\}v_\xi\, , 
\label{M_1} \\
&&\!\!\!\!\!\!\!\!\!\!\!\!\!\!\!\!\!\!\!\!
v_\xi= \left(\frac{\kappa}{\Lambda}\right)^{\overline{\beta}^2/{4\pi}}
e^{-\overline{\beta}^2/64} u_\xi = 
\left(\frac{\kappa}{\Lambda}\right)^{\overline{\beta}^2/{4\pi}}
\sqrt{\frac{\kappa}{2\pi}} e^{-\pi/32} e^{-\overline{\beta}^2/64}e^{-i\pi\xi/4}\, .
\label{M_2}
\end{eqnarray}
For the corresponding current $\widehat{J}_{(\lambda)}^\mu(x)$, defined by Eqs.
(\ref{bos-111})--(\ref{K_Z}), or by the Johnson definition \cite{Jon}, \cite{s_w},
\cite{wai}, but nevertheless with the same renormalization constant $Z_{(\lambda)}(a)$,
one finds the previous bosonization rules (\ref{nweyi19}) onto the new scalar fields
$\omega(x),\,\omega^\xi\left(x^\xi\right)$, and ${\cal W}^\xi$, 
(\ref{om_U})--(\ref{intro_008}), obeying obviously the same commutation relations 
(\ref{K_8_1})--(\ref{K_9}):
\begin{eqnarray}
\widehat{J}_{(\lambda)}^\mu(x)=\frac{1}{\sqrt{\pi}}\, \partial^\mu\omega(x),
\;\mbox{ for: }\;
Z_{(\lambda)}(a)=(\Lambda^2|a^2|)^{-{\overline{\beta}^2}/{4\pi}}.
\label{J_L}
\end{eqnarray}
Substituting the normal form (\ref{M_1}) into the solution (\ref{nweyi21}), we
immediately obtain the normal exponential of the DM for Thirring field in the form,
analogous to \cite{blot}:
\begin{eqnarray}
&&
\Psi_\xi(x)={\cal N}_\varphi
\left\{\exp\left(-i\overline{\alpha}\varphi^{-\xi}\left(x^{-\xi}\right)-
i\overline{\beta}\varphi^\xi\left(x^\xi\right)-
i\overline{\alpha}\frac{\xi}{4}{\cal Q}^\xi+
i\overline{\beta}\frac{\xi}{4}{\cal Q}^{-\xi}\right)\right\}v_\xi,
\label{nweyi29}
\end{eqnarray}
by imposing the conditions onto the parameters that are necessary to have correct
Lorentz -transformation properties corresponding to the spin $1/2$, and correct canonical
anticommutation relations (\ref{vbnmei2}), (\ref{vbnmei4}), respectively:
\begin{eqnarray}
\overline{\alpha}^2-\overline{\beta}^2= 4\pi, \quad
\overline{\beta} - \frac{\beta g}{2\pi} = 0.
\label{nweyi32}
\end{eqnarray}
Straightforward calculation of the vector current operators (\ref{bos-111})--(\ref{K_Z})
for the solution (\ref{nweyi29}) by means of Eqs. (\ref{K_9})--(\ref{nblaie19}) and
(\ref{nweyi32}), under the conditions:
\begin{eqnarray}
&&
\overline{\alpha}=\left(\frac{2\pi}{\beta}+\frac{\beta}{2}\right),\quad
\overline{\beta}=\left(\frac{2\pi}{\beta}-\frac{\beta}{2}\right),\;\mbox{ or: }\;
e^\eta=\frac{2\sqrt{\pi}}{\beta}=\sqrt{1+\frac{g}{\pi}},
\label{Kab}
\end{eqnarray}
reproduces the bosonization relations (\ref{ns94m61}), (\ref{ns94m62}), (\ref{nweyi19})
as following:
\begin{equation}
%% &&\!\!\!\!\!\!\!\!\!\!\!\!\!\!\!\!\!\!\!\!
\widehat{J}_{(\Psi)}^\mu (x)\stackrel{\rm w}{=}
\frac{\beta}{2\sqrt{\pi}}\,\widehat{J}_{(\chi)}^\mu (x)=
-\,\frac{\beta}{2\pi}\,\epsilon^{\mu\nu}\partial_\nu\phi (x),\,\mbox{ for: }\,
Z_{(\Psi)}(a)=(-\Lambda^2 a^2)^{-{\overline{\beta}^2}/{4\pi}},\;\, Z_{(\chi)}(a)=1,
\label{K_Za}
\end{equation}
demonstrating self-consistency of all the above calculations. The last equality of
Eq. (\ref{Kab}) is easily recognized as the well-known Coleman identity \cite{col}.
The weak sense of bosonization rules (\ref{K_Za}), unlike (\ref{nweyi19}), is directly
manifested by the difference of renormalization constants $Z_{(\Psi)}(a)$ and 
$Z_{(\chi)}(a)$ defined by Eqs. (\ref{J_L}), (\ref{K_Za}) for the various fields 
$\Psi(x)$ and $\chi(x)$ respectively. At the same time, by making use of (\ref{nweyi19}), 
(\ref{K_8_1}), for the Johnson commutators \cite{Jon}--\cite{wai} of Heisenberg fields 
(\ref{nweyi29}) and their currents (\ref{K_Za}):
\begin{eqnarray}
&&
\left[\widehat{J}^0_{(\Psi)}(x),\Psi(y)\right]\bigr|_{x^0=y^0}\stackrel{\rm w}{=}
- \underline{a} \Psi(y)\delta\left(x^1-y^1\right),
\label{J_0}\\
&&
\left[\widehat{J}^1_{(\Psi)} (x),\Psi(y)\right]\bigr|_{x^0=y^0}\stackrel{\rm w}{=}
-\overline{a}\gamma^5 \Psi(y)\delta\left(x^1-y^1\right),
\label{J_1}\\
&&
\left[\widehat{J}^0_{(\Psi)}(x),\widehat{J}^1_{(\Psi)}(y)\right]\bigr|_{x^0=y^0}
\stackrel{\rm w}{=}
-ic_{(\Psi)}\partial_{x^1}\delta\left(x^1-y^1\right),
\label{J_S}
\end{eqnarray}
upon the above accepted definitions and relations (\ref{Kab}) one obtains:
\begin{eqnarray}
&&
\underline{a}=1,\quad  \overline{a}=\frac{\beta^2}{4\pi}, \quad 
c_{(\Psi)}=\frac{\beta^2}{4\pi^2},
\label{J_3} \\
&&
\mbox{and finds: }\; \underline{a}\overline{a}=\pi c_{(\Psi)},\quad 
\underline{a}-\overline{a}=gc_{(\Psi)},
\label{J_33}
\end{eqnarray}
in agreement with \cite{s_w}--\cite{d_f_z}. On the other hand, in accordance with
\cite{man}, \cite{ks}, \cite{sok}, the algebra of the Heisenberg operator of the
conserved fermionic charge, by virtue of (\ref{J_0}), (\ref{J_3}), coincides with the
algebra of the  conserved fermionic charge $O/\sqrt{\pi}$ from Eq. (\ref{K_O}) for the
free trial field
$\chi(x)$. Note, that the use of relations (\ref{J_0}), (\ref{J_1}) for calculation
of the commutator in Eq. (\ref{45bn6l4}) violates the equations of motion
(\ref{45bn6l4}), (\ref{45bn6l5}), as well as the above-mentioned attempt to use the
commutator (\ref{J_S}) in equation (\ref{bnemti5}).

\section{Conclusion}

We have shown here, that the Thirring model \cite{thi}--\cite{d_f_z}, as well as the 
Federbush one \cite{ks}, is exactly solvable due to intrinsic hidden exact 
linearizability of its HEq, and that the bosonization rules make an operator sense only 
among the free fields operators. 
For the Heisenberg currents these rules are applicable only in a weak 
sense (\ref{K_Za}), that is naturally manifested also as various values of Schwinger's
terms (\ref{J_S}), (\ref{J_3}) for the free and Heisenberg currents in inequivalent field
representations (\ref{nblaie12}), (\ref{M_1}) and (\ref{nweyi29}) respectively:
\begin{eqnarray}
c_{(\chi)}=c_{(\lambda)}=\frac{1}{\pi}, \quad c_{(\Psi)}=\frac{1}{\pi+g},
\label{Sc_t}
\end{eqnarray}
in agreement with \cite{ot}. Similarly to the solution \cite{ks} of Federbush model, the
linear homogeneous HEq (\ref{nweyi13}) does not define the normalization of HF
(\ref{nweyi29}), (\ref{Kab}), which, as well as for the free fields 
$\chi(x),\,\lambda(x)$, is fixed \cite{fab-iva} only by the anticommutation
relations (\ref{vbnmei2}). We want to point out, that unlike \cite{mtu}--\cite{green} the
bosonization procedure of Refs. \cite{col}--\cite{blot} is considered
here as a particular case of dynamical mapping onto the ``Schr\"odinger'' physical
field \cite{vklt}--\cite{ks} defined at $t=0$. From this view point the results of Refs.
\cite{raja} and \cite{fab-iva} look as DM of Thirring field onto the free massive scalar
field $\phi_m(x)$, or free massive Dirac field $\psi_M(x)$ respectively. 
The general form of solution (\ref{nweyi21}) should give a possibility to describe all
phases of the theory under consideration. We postpone the discussion of these
features to subsequent works.

The authors thank A.N. Vall and V.M. Leviant for helpful discussions.

This work was supported in part by the RFBR (project N
09-02-00749) and by the program ``Development of Scientific
Potential in Higher Schools'' (project N 2.2.1.1/1483,
2.1.1/1539).

\end{document}